\documentstyle[prc,aps,manuscript]{revtex}

\begin{document}

\draft

\title{DYNAMICS AND INSTABILITIES OF NUCLEAR FERMI LIQUID}

\author{V.M.Kolomietz}

\address{Institute for Nuclear Research, 
Prosp. Nauki 47, 252028 Kiev, Ukraine}

\maketitle

\begin{abstract}
The kinetic theory of Fermi liquid is applied to finite nuclei.
The nuclear collective dynamics is treated in terms of the observable 
variables: particle density, current density, pressure etc.  
The relaxation processes and the development of instabilities in nuclear 
Fermi-liquid drop are strongly influenced  
by the Fermi-surface distortion effects.  
\end{abstract}

\vskip 2cm

\pacs{PACS numbers: 21.60.-n,21.60.Ev,24.30Cz}

\section{Introduction. Nuclear Fermi-liquid drop model}

Dynamics and dissipative properties of nuclear 
Fermi liquid depend in many aspects on the dynamic distortion of 
the Fermi surface in the momentum space. It is well-known that 
taking into account this distortion allows the description of a new class 
of phenomena, most famous of which are giant multipole resonances.
Furthermore, scattering of particles from the distorted Fermi surface
leads to relaxation of collective motion and gives  rise to fluid 
viscosity \cite{ak}. We point out that the development of instability 
in nuclear processes like binary fission or multifragmentation
in HI-reactions also depends on dynamic effects such as the 
dynamic Fermi-surface distortion.

A convenient way to introduce the Fermi-liquid effects into 
the nuclear many body problem is to start from the kinetic equation
for the phase space distribution function $f (\vec{r}, \vec{p}, t)$.
The kinetic equation can then be transformed identically to a set 
(infinite) of equations for the moments of $f (\vec{r}, \vec{p}, t)$ in 
${\vec p}$-space, namely, local single-particle density $\rho$, the velocity
field $\vec{u}$, the pressure tensor $\pi_{\alpha \beta}$, etc.,
see \cite{ak}. In case of small variations of the particle
density, $\delta \rho$, the first order moment of the kinetic equation 
has the form of Euler-Navier-Stokes equation and is given by
\cite{kt}
\begin{equation}
m \rho_{eq} {\partial \over\partial t} u_{\alpha} +
\rho_{eq} { \partial \over\partial r_{\alpha}}
\left({\delta^{2} {\cal E} \over \delta \rho^{2}}\right)_{eq}
\delta\rho + { \partial \over\partial r_{\nu}}\pi_{\nu\alpha} = 0.
\label{eiler}
\end{equation}
The internal energy density ${\cal E}$ in Eq. (\ref{eiler}) contains
both kinetic ${\cal E}_{kin}$ and potential ${\cal E}_{pot}$ 
energy densities: $ {\cal E}={\cal E}_{kin}+{\cal E}_{pot}$.
The pressure tensor $\pi_{\alpha \nu}$ depends on the Fermi-surface
distortion effect. In general case, tensor $\pi_{\alpha \nu}$
also includes the viscosity tensor which is derived by the
collision integral.

Eq. (\ref{eiler}) is not closed because it contains the pressure
tensor $\pi_{\alpha \beta}$ given by the second order moment of 
the distribution function $f (\vec{r}, \vec{p}, t)$.
We will follow the nuclear fluid dynamic approach 
\cite{Kob,HoEc1,KoPlSh1} and take into account dynamic 
Fermi-surface distortions up to the multipolarity of $l = 2$. Assuming
a periodic in time eigenvibrations with the eigenfrequency $\omega =
\omega_0 + i\Gamma/2\hbar$, where $\omega_0$ and $\Gamma$ are real,
and a separable form of the velocity field,
$\vec{u}(\vec{r},t) = \beta(t) \vec{v}(\vec{r})$,
with $\beta (t) \sim e^{-i\omega t}$,
Eq.  (\ref{eiler}) is reduced to the equation of motion
for the macroscopic variable $\beta (t)$ with the following
secular equation 
\begin{equation}
-\omega^2 B + (C^{(LD)} + 
C^\prime (\omega )) - i\omega \gamma (\omega) = 0.
\label{6.7}
\end{equation}
Here, $B$ and $C^{(LD)}$ are the mass coefficient
and the stiffness coefficient in the traditional liquid drop
model (LDM) \cite{BoMo} respectively.
The additional contribution from $C^{\prime}(\omega)$
to the stiffness coefficient 
and the  dissipative term $\gamma(\omega)$ depend on the
relaxation time $\tau$ and are given by
\begin{equation}
C^{\prime}(\omega)=\displaystyle \int d\vec{r} \,\,P_{eq}{(\omega_{0}
 \tau)^2 \over 1+(\omega_{0}\tau)^{2}}\,
\Big({\partial v_{\alpha} \over\partial r_{\nu}} +
{\partial v_{\nu} \over\partial r_{\alpha}}
- {2\over 3}\,\delta_{\alpha\nu}\,
{\partial v_{\lambda} \over\partial r_{\lambda}}\Big)\,
 {\partial v_{\alpha} \over\partial r_{\nu}}
\label{cc}
\end{equation}
and
\begin{equation}
\gamma(\omega)= \int d\vec{r}\,\,P_{eq}{\tau \over
1+(\omega_{0} \tau)^{2}}\,
\Big({\partial v_{\alpha} \over\partial r_{\nu}} +
{\partial v_{\nu} \over\partial r_{\alpha}}
- {2\over 3}\,\delta_{\alpha\nu}\,
{\partial v_{\lambda} \over\partial r_{\lambda}}\Big)\,
{\partial v_{\alpha} \over\partial r_{\nu}}, 
\label{gamma}
\end{equation}
where $P_{eq}$ is the equilibrium pressure of the Fermi gas. 
The additional contribution from
$C^{\prime}(\omega)$ to the stiffness coefficient
in Eq. (\ref{6.7}) is absent in the LDM, i.e.
in the liquid drop limit $\omega\tau \to 0$, and 
represents the influence of the dynamic Fermi-surface distortion
on the conservative forces in the Fermi system. 
Finally, the dissipative term $\gamma(\omega)$ 
appears due to the interparticle scattering from the
distorted Fermi surface.

In general, both, $C^\prime(\omega )$ and $\gamma (\omega )$ 
depend implicitly on the temperature, $T$, via the dependence of the 
relaxation time $\tau$ and of $P_{eq}$ on $T$.
In cold nuclei, in the zero-sound limit $\omega \tau \to \infty$,
the main contribution to the stiffness coefficient in Eq. (\ref{6.7})
is due to the Fermi-surface distortion effect given by $C^\prime (\omega )$.
In Fig. 1, this effect is shown in a transparent way 
for isoscalar quadrupole excitations. As it is seen from this figure,
the Fermi-surface distortion effect
leads to a significant upward shift of the energy of vibrational
states to locate it in the region of the quadrupole giant
resonance (solid line).

\section{Giant monopole resonance and nuclear incompressibility}

We will discuss the Fermi-surface distortion effect in more detail 
for the case of isoscalar giant monopole resonances (ISGMR).
This particular case is important for understanding the
nature of nuclear incompressibility. We will consider below the
model for a Fermi-liquid drop having a sharp surface of the equilibrium
radius $R_0$ and the bulk density $\rho_0$.

The particle density variation $\delta \rho$ is then given by
\begin{equation}
\delta \rho (r, t) = \eta (r, t)\,\rho_0\,\theta (R_{0} - r) +
q (t)\,\rho_0 \,R_{0} \delta (R_{0} - r),
\label{drho2}
\end{equation}
where the unknown functions $\eta (r, t)$ and $q (t)$ are related to
each other by the particle number conservation and the bulk
density parameter $\eta (r, t)$ is found from the equation 
of motion derived by  Eq. (\ref{eiler}). Namely,
\begin{equation}
m\,{\partial^2\over \partial t^2}\,\eta = {1\over 9}\,K^\prime \,
\nabla^2 \eta\,\,\,\,\,\,\,\,\,{\rm with}\,\,\,\,\,\,\,\,\,
K^\prime = K + K_\mu .
\label{eq2}
\end{equation}
Here $K$ is the {\it static} incompressibility
\begin{equation}
K = R^2\,{\delta^2 E/A\over \delta R^2}\Big\vert_{R=R_{0}}.
\label{kstatic}
\end{equation}
The additional contribution $K_\mu$ to the incompressibility
$K^\prime$ in Eq. (\ref{eq2}) is due to the {\it dynamic} Fermi-surface
distortion effect \cite{Kob,KoMaPl1}. The value of $K_\mu$ depends
on the Landau scattering amplitude $F_0$. In nuclear case,
$F_0 \sim 0$, one has $K_\mu \approx 2\,K$.

An essential property of a finite liquid drop having a free surface
is that the motion of the surface should be consistent with the
motion of the liquid inside the drop. This can be achieved by 
imposing a boundary condition for the compensation
of the compressional pressure $\pi_{rr}$ at the liquid surface 
by the pressure generated by the surface tension forces
$\delta P_\sigma$. Finally, the eigenenergies in Eq. (\ref{eq2})
are given by
\begin{equation}
\hbar \,\omega_n = \sqrt{{\hbar^2\,K^\prime \over 9\,m\,R_{0}^2}}\,x_n,
\label{omn}
\end{equation}
where $x_n = k_n\,R_0$ are derived from the following boundary
condition:
\begin{equation}
x_n\,j_0 (x_n) - (f_\sigma + f_\mu )\,j_1 (x_n) = 0.
\label{sec}
\end{equation}
Here the coefficients $f_\sigma$ and
$f_\mu$ are related to the surface tension and the Fermi-surface
distortion respectively and are given by
\begin{equation}
f_\sigma = {18\,\sigma \over \rho_0\,R_0\,K^\prime},
\,\,\,\,\,\,\,\,\,\,\,
f_\mu = {3\,K_\mu \over K^\prime}.
\label{ff}
\end{equation}
In the general case of Fermi-liquid drop with $f_\mu \neq 0$,
the eigenfrequency $\omega_n$ given in Eq. (\ref{omn}) is renormalized
due to two contributions associated with the Fermi-surface distortion:
1) the direct change of the sound velocity, i.e. in Eq. (\ref{omn})
$K^\prime$ appears instead of $K$;
2) the change of the roots $x_n$ of the secular equation (\ref{sec})
due to additional contribution from $f_\mu \neq 0$ in 
Eq. (\ref{sec}). These two effects work in opposite directions:
$K^\prime$ increases $\omega_n$ while $f_\mu$ in 
Eq. (\ref{sec}) decreases it.
Both values of $K^\prime$ and $x_n$ in Eq. (\ref{omn}) depend on the 
Landau scattering amplitude $F_0$. 
In the limit of zero- to first sound transition at $F_0 \gg 1$,
one has $f_\mu \to 0$ and the solution of Eqs. (\ref{sec}) and
(\ref{omn}) is close do the one obtained in LDM.
However, for realistic nuclear forces  $F_0 \sim 0$, the general
expressions (\ref{sec})-(\ref{omn}) with $f_\mu \neq 0$
have to be used.

In Fig. 2 we plotted the eigenenergy, $E_{0^+}$,
of the lowest monopole mode $(n = 1)$ given by Eqs. (\ref{omn}) and 
(\ref{sec}) as a function of mass number $A$. The dashed line shows
the result for $E_{0^+}$ from Eq. (\ref{omn}) where $K^\prime$
was replaced by $K_{LDM} = 190 \,MeV$ and 
boundary condition (\ref{sec}) with $f_\mu = f_\sigma =0$ was
used. The solid line shows the result of FLDM calculations.
We adopted the following expression for $K$  
\cite{BlGr} $K = K_\infty + K_\sigma\,A^{-1/3}$ with 
$K_\infty = 220\,MeV$ and $K_\sigma = -440\,MeV$ and for
$\rho_0 = 0.17\,fm^{-3},\,\,\, \epsilon_F = 40\,MeV,\,\,\,
r_0 = 1.12\,fm$ and $\sigma = 1.2\,MeV/fm^2$. 
So, in spite of strong renormalization of nuclear
incompressibility due to the Fermi-surface distortion effect,
the energy of the giant monopole resonance $E_{0^+}$ in
the Fermi-liquid drop is located close to the one in the traditional 
LDM \cite{BoMo} where the Fermi-surface distortion effects are 
absent. The compensation of the Fermi-surface distortion
effect in nuclear incompressibility $K^\prime$ due to
the consistent account of the same effect in the boundary condition  
(\ref{sec}) is essential for the lowest mode
$(n= 1)$ only. The above mentioned compensation is rather small for the
highest modes with $n \geq 2$. In particular,
the eigenenergy of the double (overtone) 
giant monopole resonance $E_{0_2^+}$ (in contrast to $E_{0^+}$) 
has a significant upward shift with respect to  
$E_{0_2^+}^{LDM} \approx 2\,E_{0^+}$ given by the traditional 
liquid drop model. 

\section{Collisional relaxation on the deformed Fermi 
surface}

The relaxation of nuclear collective motion toward thermal equilibrium
has been described in great detail within the framework of the
kinetic theory, taking into account the collision integral 
$\delta St (\vec{r},\vec{p},t)$
\cite{Kob,KoPlSh1,KoMaPl1,SaYo,KoPlSh2,KoLuPlSh}. 
The collision integral $\delta St (\vec{r},\vec{p},t)$
depends on the equilibrium distribution function,
$f_{eq}(\vec{r}, \vec{p})$. In case of a finite Fermi system,
the equilibrium distribution function, $f_{eq}(\vec{r}, \vec{p})$,
contains both the diffusive layer and oscillations in $\vec{p}$-space 
which are caused by particle reflections at the potential surface. Both of 
them depend on the distance $r$. This fact also leads to the 
$r$-dependence of the collisional relaxation time, $\tau_2(r, \omega)$,
given by
\begin{equation}
1/\tau_{2}(r, \omega ) = -\int d\vec{p}\, p^2\, Y_{20} \,\delta St/ 
\int d\vec{p}\, p^2\, Y_{20}\, \delta f.
\label{tau2}
\end{equation} 
The $\omega$-dependence of $\tau_{2}(r, \omega )$
is due to the memory effects in the collision integral \cite{KoMaPl1}.

The evaluation of the collisional relaxation time $\tau_2(r, \omega)$ 
in a finite Fermi system requires additional accuracy. The   
diffusive layer in the equilibrium distribution function in momentum
space can lead to a spurious effect in non-zero gain- and 
loss fluxes of the probability in the ground state of the Fermi system. 
However, it can be  shown, see Ref. \cite{KoLuPlSh}, that the general
condition for the disappearance of both above mentioned
fluxes is reached due to the occurrence of quantum oscillations 
in the equilibrium distribution function $f_{eq}(\vec{r}, \vec{p})$
in momentum space.
The $r$-dependent diffusive tail of the distribution
function in momentum space leads to an increase of the collisional 
damping of the collective motion in the surface region of a nucleus
and thus to an increase of the isoscalar giant quadrupole resonance
(GQR) width. However, this increase 
is strongly reduced  due to the above mentioned oscillations of 
the equilibrium distribution functions appearing in the collision
integral. As a result, the collisional width of the
isoscalar GQR does not exceed 30-50\% of the experimental value
and agrees with the estimates of the width where the sharp
Thomas-Fermi distribution function is used \cite{KoLuPlSh}. 

In the case of heated nuclei, both temperature and memory
effects have to be taken into account in the collision integral.
For frequencies small compared to the Fermi energy, one finds 
\cite{KoMaPl1,ando}:
\begin{equation}
1/\tau_{2} =
[1+ (\hbar \,\omega _{0}/2\,\pi \,T)^{2}]/ \tilde{\tau}_{2},
\label{3.20}
\end{equation}
which is valid for $T,\,\hbar \omega _{0} \ll \epsilon_F$
and when sharp Thomas-Fermi distribution for $f_{eq}(\vec{r}, \vec{p})$
is assumed.
The magnitude of $\tilde{\tau}_{2}$  can be 
given in the following general form at low temperatures  
$T \ll \epsilon_F$ \cite{KoPlSh1}
\begin{equation}\label{3.22}
\tilde{\tau}_{2}(T)/\hbar  =\alpha^{(\pm )}  T^{-2} ,
\end{equation}
where the quantity $\alpha^{(+)}$ is for the isoscalar mode and 
$\alpha^{(-)}$ is for the isovector mode:
$\alpha^{(+)} = 9.2 \,MeV \,\,\,\,\, {\rm and} \,\,\,\,\,\, 
\alpha^{(-)} = 4.6 \,MeV$ \cite{KoPlSh2}. 

In general, there are several contributions to the relaxation
time $\tau$. The main contribution $\tau_2$ arises from the above 
discussed interparticle collisions  and the other from collisions of 
nucleons with the moving nuclear surface (one-body dissipation with 
relaxation time $\tau_1$). We will also include the contribution from
particle emission with relaxation time $\tau_{\uparrow}$. Thus,
\begin{equation}
1 / \tau = 1 / \tau_{1}
+ 1 / \tau_{2} + 1/ \tau_{\uparrow}.
\label{3.12}
\end{equation}
Introduction of the one-body dissipation reflects the peculiarities 
of our consideration, namely, restricting Fermi-surface distortions 
to the multipolarity $l \leq 2$  does not allow us to 
take into consideration Landau damping (fragmentation width). 

The total width, $\Gamma$, of a collective state can be derived
from Eq. (\ref{gamma})  with $\tau_2$ replaced by the total relaxation
time $\tau$ from Eq. (\ref{3.12}). Assuming sharp nuclear surface,
the total width $\Gamma$ is reduced to the following form,
see Ref. \cite{KoPlSh2},
\begin{equation}
\Gamma \approx  2\,x\,\hbar \omega _{0}\,
{{\omega _{0}\tau}\over {1 + x\,{(\omega _{0}\tau)}^{2}}} ,
\label{3.10}
\end{equation}
where $x \approx (1/6)\,(v_F/u_1)^2$ and $u_1$ is the first 
sound velocity in nuclear matter.
We want to note that the total intrinsic width given by Eq. 
(\ref{3.10}) has a bell-shaped form as a function of $y =  
\omega_0 \tau $. The  width $\Gamma $ is peaked around $y \equiv y_0 =
x^{-1/2}$ and the maximum value of $\Gamma$ is $\Gamma _{max}
= \hbar \omega_0 x^{1/2}$. It is easy to see that $y_0$ represents the
crossing point of both curves $\Gamma^{(0)}(y) \sim 1/\tau$ (zero sound
regime) and $\Gamma^{(1)}(y) \sim \tau$ (first sound regime). 
Due to this fact the condition
\begin{equation}
\omega_0 \tau \equiv y_0 = (\omega_0 \tau)_0  = x^{-1/2}
\label{4.4}
\end{equation}
can be used as the condition for the transition from long to 
short relaxation time regimes. The magnitude of the intrinsic 
width $\Gamma $ decreases when the parameter $\omega_0 \tau$ exceeds 
$y_0$. We have $y_0 = 2.28$ for the 
isovector GDR at the realistic value of $x = 0.19$. Such a value 
of $\omega_0 \tau$ can be reached 
at the temperature $T \equiv T_{\rm tr} \approx 4.5 \,MeV$, see 
Eq. (\ref{3.20}).  If the relation (\ref{4.4}) with $x=1$ 
and $\tau =\tau_2$ is used as the condition for transition 
between different regimes of sound wave propagation  
then the transition between both regimes occurs at the higher 
temperature $T_{\rm tr} \approx 10 \,MeV$. 
In Fig. 3 we show the intrinsic width of the giant 
dipole resonance  (GDR) in the nucleus $^{112}${\it Sn}. 
As it is seen from Fig. 3, the expression 
(\ref{3.10}) (solid line) leads to a smoother behaviour of the 
total intrinsic width with increasing excitation energy as compared to 
the prediction of the  zero-sound regime given by 
the following condition:  $\Gamma \approx 2\hbar/\tau = 
\Gamma_1 + \Gamma_2 + \Gamma_{\uparrow}$ 
with $\Gamma_1 = 2\hbar/\tau_1,\,\,\,
\Gamma_2 = 2\hbar/\tau_2$ and $\Gamma_{\uparrow} 
= 2\hbar/\tau_{\uparrow}$ (dashed line).  
Our calculation of the intrinsic width $\Gamma$
for the isovector GDR in the hot nucleus $^{112}${\it Sn} confirms a
saturation effect in the energy dependence of $\Gamma$. However,
we observe a systematic deviation of the evaluated width with respect
to experimental values. This deviation can be reduced by varying 
the parameter $x$ in Eq. (\ref{3.10}). An increase in the  
parameter $x$ improves agreement with the experiment (see dotted line 
in Fig. 3).

\section{Bulk instability}

In the vicinity of the equilibrium state of the nuclear Fermi-liquid
drop the stiffness coefficients for particle density and surface
changes are positive and the nucleus is
stable with respect to the corresponding distortions. With 
decreasing bulk density or increasing internal excitation energy
(temperature) the liquid drop reaches regions of mechanical
or thermodynamical instabilities with respect to the small particle
density and experiences shape fluctuations and separation into liquid and
gas phases. 

Let us consider small density fluctuations $\delta \rho ({\bf r}, t)$
starting from Eq. (\ref{eiler}). 
We will use a simple saturated equation of state ${\cal E} \equiv
{\cal E}[\rho ] = {\cal E}_{kin}[\rho ] + {\cal E}_{pot}[\rho ]$
with
$$
{\cal E}_{kin}[\rho ] = {\hbar^2\over 2\,m}\,
\left[{3\over 5}\,\Big({3\pi^2\over 2}\Big)^{2/3} \,\rho^{5/3} +
{1\over 4}\,\eta \,{(\vec{\nabla}\rho)^2\over \rho}\right],\,\,\,
$$
\begin{equation}
\label{epot}
\end{equation}
$$
{\cal E}_{pot}[\rho ] = t_0\,\rho^2 + t_3\,\rho^3 +
t_s\,(\vec{\nabla}\rho)^2 + \epsilon_C[\rho ].
$$
The potential energy ${\cal E}_{pot}[\rho ]$ has a Skyrme-type form
including the surface term $\sim (\vec{\nabla}\rho)^2$
and the Coulomb energy density, $\epsilon_C[\rho ]$. 

We will again assume a sharp surface behaviour of the equilibrium
particle density.
Assuming also $\delta \rho \sim \exp (-i\omega t)$ and
taking into account continuity equation  and Eq.
(\ref{epot}), the equation of motion (\ref{eiler}) can 
be reduced to the following form (we consider the isoscalar mode):
\begin{equation}
- m\,\omega^2 \,\delta\rho = \left({1\over 9}\,K -
{4\over 3}{i\omega \tau  
\over {1 - i\omega \tau}} (P_0/\rho_0)\right)
\nabla^2\delta \rho 
- 2\,(\beta + t_s\,\rho_0)\,\nabla^2 \nabla^2 \delta\rho,
\label{0eq3}
\end{equation}
where $P_0 = \rho_0\,p_F^2/5\,m$, $\,\,\,\beta = (\hbar^2/ 8\,m)\,\eta$
and $K$ is the incompressibility
coefficient
\begin{equation}
K \approx 6\,\epsilon_F\,(1 + F_0)\,\,\,\,
{\rm with}\,\,\,\,F_0 = {3\,\rho_0\over \epsilon_F}\,(t_0 + 3\,t_3\,\rho_0).
\label{K1}
\end{equation}

Let us  consider volume instability regime, $K < 0$, and
introduce the growth rate $\Gamma = -i\,\omega$ ($\Gamma$ is real,
$\Gamma > 0$). Using Eq. (\ref{0eq3}), one obtains
\begin{equation}
\Gamma^2 = |u_1|^2 \,q^2 - \zeta (\Gamma )\,q^2 - \kappa_s\,q^4,
\label{0disp}
\end{equation}
where
\begin{equation}
u_1^2 = (1/9m)\,K,\,\,\,\,
\zeta (\Gamma ) = {4\over 3\,m}{\Gamma \tau \over
{1 + \Gamma \tau}}{P_{0}\over \rho_{0}},
\,\,\,\,\kappa_s =(2/m)(\beta + t_s\,\rho_0).
\label{zeta}
\end{equation}
The quantity $\zeta (\Gamma )$ in Eq. (\ref{0disp}) appears due
to the Fermi surface distortion effect.
Equation (\ref{0disp}) is valid for arbitrary relaxation times $\tau$ 
and thus describes both rare and 
frequent collision limits as well 
as intermediate cases. From it one can obtain the leading order
terms in the different limits mentioned. 

{\it (i) Frequent collision regime: $\tau \to 0$.
High temperatures.}\\
The contribution from the dynamic distortion of the Fermi surface,
$\kappa_v$, can be neglected in this case and we have from Eqs.
(\ref{0disp}) and (\ref{gamma}),
\begin{equation}
\Gamma^2 = |u_1|^2 \,q^2 - \Gamma\,(\gamma /m)\,q^2 - \kappa_s\,q^4,
\label{0disp1}
\end{equation}
where $\gamma = (4/3)\tau\,P_0/\rho_0$ is the classical friction
coefficient. In the limit of small friction, we obtain
\begin{equation}
\Gamma^2 \approx |u_1|^2\,q^2 - \kappa_s\,q^4 -
{\gamma \over m}\,q^2\,\sqrt{|u_1|^2\,q^2 - \kappa_s\,q^4}
\label{0disp2}
\end{equation}

The amplitude of the density oscillations with a certain
multipolarity $L$, $\delta \rho_L({\bf r}, t)$,
grows exponentially if $\Gamma > 0$. The expression (\ref{0disp2})
determines two characteristic values of the wave number $q$, namely,
$q_{max}$ and $q_{crit}$ where the growth rate reaches
the maximum of magnitude $\Gamma_{max}$ and $\Gamma$ goes to zero
(excluding $q = 0$) respectively, i.e.,
\begin{equation}
\Gamma = \Gamma_{max}\,\,\,\, {\rm at}\,\,\,\, 
q = q_{max} < q_{crit},\,\,\,\,
{\rm and}\,\,\,\,\Gamma = 0\,\,\,\,{\rm at}\,\,\,\, q = q_{crit}.
\label{def1}
\end{equation}
Both $q_{max}$ and $q_{crit}$ are given by
\begin{equation}
q_{crit}^2 = {|u_1|^2\over \kappa_s^\prime},
\,\,\,\,\kappa_s^\prime = \kappa_s + {1\over m^2}\,\gamma^2,\,\,\,\,\,\,
{\partial \Gamma\over \partial q}\Big\vert_{q=q_{max}}
= 0. 
\label{crit}
\end{equation}

For a saturated nuclear liquid, one has $t_0 < 0,\,\,\, t_3 > 0$
and $t_s > 0$. Thus, the critical value $q_{crit}$ from
Eq. (\ref{crit}) increases with a decrease of the bulk density
$\rho_0$, see 
Eqs. (\ref{K1}) and (\ref{crit}). The presence of friction
decreases the critical value $q_{crit}$, i.e. reduces the 
instability. The existence of the critical wave 
number $q_{crit}$ for an unstable mode is a feature of a finite 
system. The growth rate $\Gamma$ depends on the multipolarity 
$L$ of the nuclear density distortion and on the position of $q_L$ 
in the interval $q = 0\,\div\,q_{crit}$ \cite{PeRa2}. 
The value of $q_L$ is determined from the boundary condition for
the pressure on free nuclear surface. For
given $R_0$ the value of $q_L$ increases with $L$ at $L \geq 2$.
That means that
if $q_L < q_{max}$ instability increases with $L$ and the 
nucleus becomes more unstable with respect to internal 
clusterization into small pieces (high multipolarity
or multifragmentation regimes)
rather than to binary fission (low multipolarity regime). In
contrast, the binary fission is preferable if $q_{max} < q_L <
q_{crit}$.

{\it (ii) Rare collision regime: $\tau \to \infty$.
Cold expansion.}\\
Neglecting the contribution from the relaxation, 
one has from Eq. (\ref{0disp})
\begin{equation}
\Gamma^2 = |u_1|^2\,q^2 - \kappa_v^\prime \,q^2 -\kappa_s\,q^4,
\,\,\,\,\,\,\kappa_v^\prime = {4\over 3\,m}{P_{eq}\over \rho_{eq}}.
\label{0disp3}
\end{equation}

The critical value $q_{crit}$ and the value $q_{max}$ are given by
\begin{equation}
q_{crit}^2 = {{|u_1|^2 - \kappa_v^\prime}\over \kappa_s},
\,\,\,\,\,\,\,\,\,\,\,\, q_{max}^2 = {1\over 2}\,q_{crit}^2.
\label{crit1}
\end{equation}
Thus, the distortion of the Fermi-surface leads to a decrease
of the critical value $q_{crit}$ i.e. the Fermi-liquid
drop becomes more stable with respect to volume density
fluctuations due to dynamic Fermi-surface distortion effects.

\newpage

\begin{figure}
\caption{Energies of the lowest isoscalar $2^+$ excitations
as evaluated in the traditional liquid drop model [6]
(LDM), i.e., at $C^\prime (\omega ) = 0$ 
and in the Fermi-liquid drop model (FLDM) taking into account
both contributions $C^{(LD)}$ and $C^\prime (\omega )$ in 
Eq. (2). The experimental data are from Ref. [15].}
\end{figure}

\begin{figure}
\caption{Energies of the isoscalar monopole resonances
in the liquid drop model (dashed line) and in the Fermi-liquid drop
model (solid line). Experimental data from Ref. [15].}
\end{figure}

\begin{figure}
\caption{The intrinsic width of the isovector 
giant dipole resonance in the nucleus $^{112}Sn$ 
as function of  excitation  energy $U$. The experimental data were 
taken from Ref. [16]. Solid line corresponds to the total 
width.  Dashed line is the width in the zero sound approximation 
at $\Gamma = 2\hbar/\tau$. Dotted line is the fit to the experimental
data due to the parameter $x=0.78$ in Eq. (15), see Ref. [5].}
\end{figure}

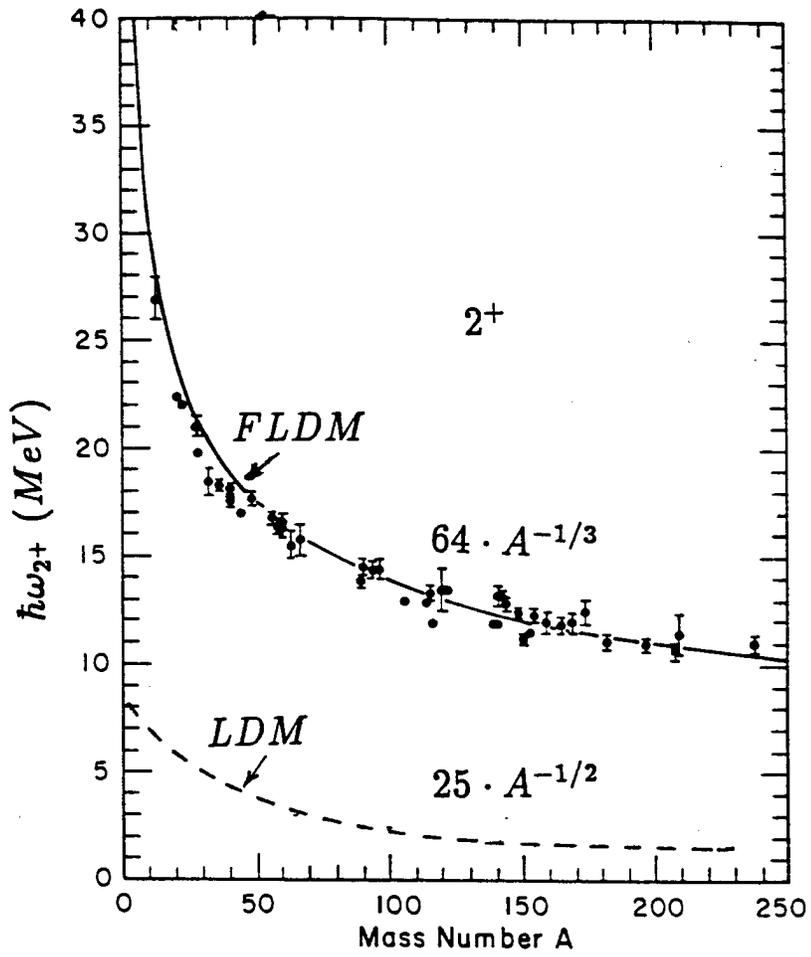
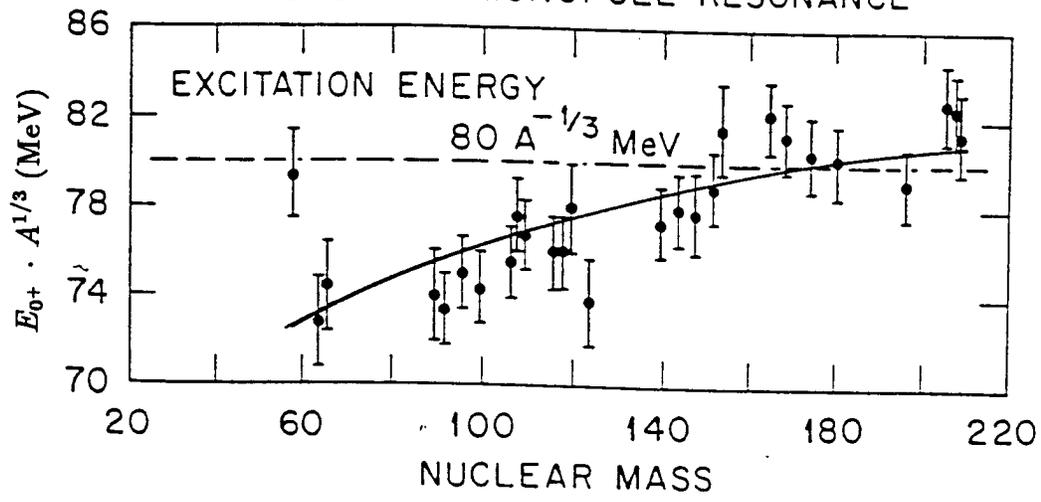
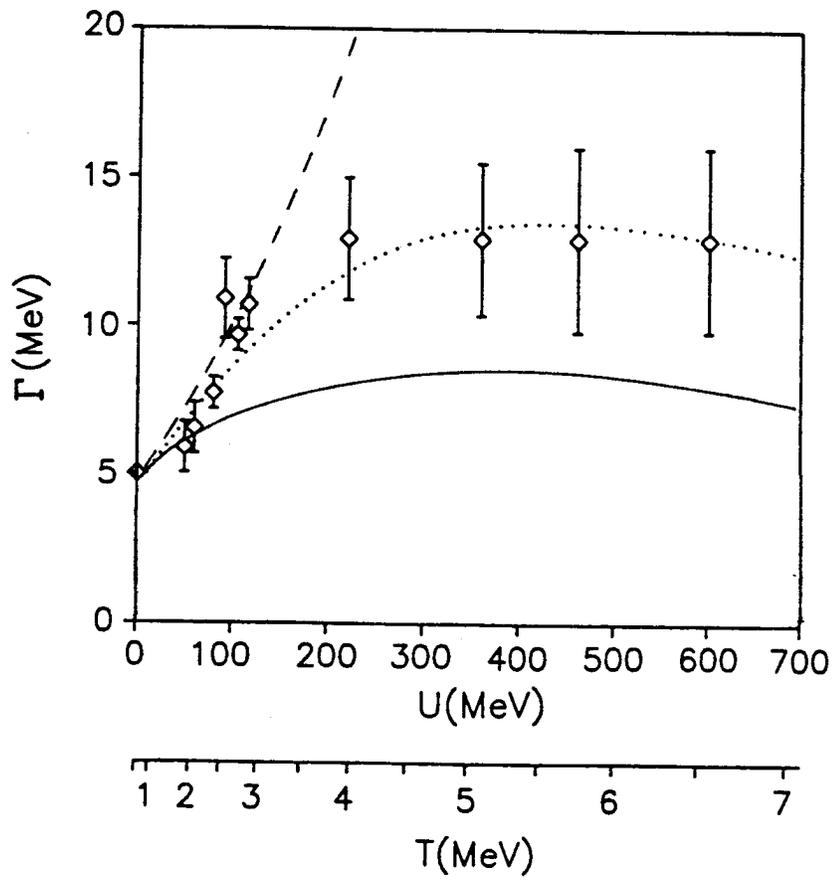
\end{document}